\documentclass[prl,twocolumn,showpacs]{revtex4}
\usepackage{amsfonts,amsmath,mathrsfs,epsfig,amsbsy,bm,verbatim}

\begin{document}

\title{Disentangling Majorana fermions from topologically trivial low--energy states in semiconductor Majorana wires}
\author{T. D. Stanescu$^1$}
\author{Sumanta Tewari$^{2}$}

\affiliation{$^1$Department of Physics, West Virginia University, Morgantown, WV 26506\\
$^2$Department of Physics and Astronomy, Clemson University, Clemson, SC
29634}

\begin{abstract}
Majorana fermions (MFs) are predicted to occur as zero--energy bound states in semiconductor nanowire--superconductor structures. However, in the presence of disorder or smooth confining potentials, these structures can also host  non--topological nearly--zero energy states. Here, we demonstrate that the MFs and the nearly--zero topologically--trivial states have different characteristic signatures in a tunneling conductance measurement, which allows to clearly discriminate between them. We also show that low--energy non--topological states can strongly hybridize with metallic states from the leads, which generates the smooth background that characterizes the soft superconducting gap measured in tunneling experiments and produces an additional decoherence mechanism for the Majorana mode. Our results pave the way for the conclusive identification of MFs in a solid state system and provide directions for minimizing quantum decoherence in Majorana wires.
\end{abstract}

\maketitle

Majorana fermions (MFs) \cite{Majorana} -- quantum (quasi) particles representing their own anti-particles \cite{Majorana,Wilczek,Frantz} first introduced as possible purely real solutions of the Dirac equation \cite{Majorana} -- have been proposed to exist in low temperature systems in the context of the fractional quantum Hall effect \cite{Nayak-Wilczek,Read-Green},  chiral $p$-wave superconductors/superfluids \cite{Read-Green},   topological insulator heterostructures \cite{Fu-Kane}, and cold fermion systems \cite{Zhang-Tewari,Sato-Fujimoto}. More recently, it has been shown that spin-orbit coupled semiconductor thin films \cite{Sau,Long-PRB} and nanowires \cite{Long-PRB,Roman,Oreg}  with Zeeman spin splitting and proximity induced $s$-wave superconductivity can also host MFs as zero energy bound states.
 The 1D version -- the so-called semiconductor Majorana wire -- is a direct physical realization of the Kitaev model \cite{Kitaev-1D} and has recently received considerable experimental attention \cite{Mourik,Deng,Weizman,Rokhinson}. For small Zeeman splitting $\Gamma$, the system is in a conventional (proximity-induced) superconducting (SC) state with no MFs, while for $\Gamma$ larger than a critical value $\Gamma_c$, localized MFs exist at the wire ends.  A zero-energy MF can be detected in charge tunneling measurements~\cite{Long-PRB,Sengupta-2001,bolech,R1} at experimentally accessible temperatures as a sharp zero bias conductance peak (ZBCP). 
 
Despite its apparent conceptual simplicity, the ZBCP experiment does not constitute a sufficient proof for the existence of MFs in Majorana wires. A nearly zero bias peak can occur even in the topologically trivial phase, i.e.,  in the absence of MFs,  when the confinement potential at the wire ends is smooth \cite{Kells}, or in the presence of strong disorder\cite{Liu}.
Therefore, it is critical to identify a diagnostic signature that allows to clearly distinguish between a ZBCP arising from MFs and these more conventional nearly--ZBCPs that may appear in the topologically trivial phase. More generally, it is paramount to determine the mechanisms responsible for the occurrence of non--MF low--energy in--gap gap states and establish the role of these states in the quantum decoherence of the Majorana mode. 
Here, we show that, for a wire with strong disorder or smooth end--of--wire confinement, the emergence of the nearly--ZBCPs is necessarily accompanied by a signature in the end--of--wire local density of state (LDOS) similar to the closing of a gap (henceforth, referred to as the ``gap closing signature''). We emphasize that this is associated with in--gap states, while the bulk gap remains finite, hence there is no corresponding quantum phase transition. In contrast, a ZBCP that occurs without a gap closing signature is due to the presence of MFs localized at the end of the wire. We also find that the characteristic soft superconducting gap observed experimentally \cite{Mourik,Deng,Weizman} is due to in--gap states associated with lower energy bands that hybridize strongly (much stronger that MFs) with metallic states from the leads.
   
We consider a rectangular semiconductor (SM) nanowire with dimensions $L_x\gg L_y \sim L_z$ proximity coupled to an $s$-wave superconductor.
For an infinite  wire, $L_x\rightarrow \infty$, the effective BdG Hamiltonian has the form,
\begin{eqnarray}
H_{nm}(k) &=&[\epsilon_{nm}(k) -\mu \delta_{nm}]\tau_z + \Gamma\delta_{nm}\sigma_x\tau_z \nonumber \\
&+& \alpha k \delta_{nm} \sigma_y\tau_z - i \alpha_y q_{nm} \sigma_x +\Delta_{nm}\sigma_y\tau_y,      \label{Hnm}
\end{eqnarray}
where $k=k_x$ is the wave number, $\sigma_i$ and $\tau_i$ are Pauli matrices associated with the spin-1/2 and the particle-hole (p--h) degree of freedom, respectively, and we use the basis $(u_{\uparrow}, u_{\downarrow}, v_{\uparrow}, v_{\downarrow})$ for the p--h spinors.  In Eq.~(\ref{Hnm}) $n =(n_y,n_z)$ and  $m=(m_y,m_z)$ label different confinement-induced bands described by the transverse wave functions $\phi_n(y) \propto \sin( n_y\pi y/L_y)\sin( n_z\pi z/L_z)$, $\epsilon_{nm}$ describes the SM spectrum without spin--orbit coupling re--normalized by proximity effect, $\mu$ is the chemical potential, $\Gamma= g^* \mu_B B/2$ is the external Zeeman field along the $x$-direction, $q_{nm}=-q_{mn}$ represent inter--band spin--orbit coupling matrix elements, and $\Delta_{nm}$ is the proximity--induced pair potential.  
The Rashba coupling is $\alpha=a\alpha_y=0.2$ eV\AA, with $a\approx 6.5$\AA ~being the lattice constant, and the parameters $\epsilon_{nm}$, $q_{nm}$, $\Delta_{nm}$ are calculated numerically following the procedure described in Ref.~[\onlinecite{SLDS}]. Throughout this paper by ``band" we mean a pair of spin sub--bands that are
degenerate at all $k_x$ in the absence of SO coupling and Zeeman field. The chemical potential $\mu$ is measured relative to the energy of the top occupied band at $k_x=0$ and $B=0$.

With increasing $\Gamma$, the SM wire undergoes a topological quantum phase transition (TQPT) at $\Gamma=\Gamma_c$.
For $\Gamma > \Gamma_c$, the presence of MFs  can be revealed as a sharp ZBCP when charge current is tunneled into  the end of the wire \cite{Long-PRB}. 
The SC  quasiparticle gap {\em must} vanish at the TQPT \cite{Long-PRB,Read-Green,SLDS}. This closing of the bulk gap is visible in the total DOS, but may have no signature in the end-of wire LDOS \cite{Stanescu-Tewari}. For $\Gamma < \Gamma_c(\mu)$,  the system is topologically trivial and there are no zero energy  MFs. However, it has been shown recently \cite{Kells} that, even in this topologically trivial phase, near-zero-energy states with a significant spectral weight near the wire ends are possible in the presence of a soft (rather than hard-wall) confinement potential. These states generate nearly-ZBCPs similar to those associated with Majorana physics. Here we identify a key qualitative feature that allows one to discriminate between these ZBCPs of different origins.

\begin{figure}[tbp]
\begin{center}
\includegraphics[width=0.5\textwidth]{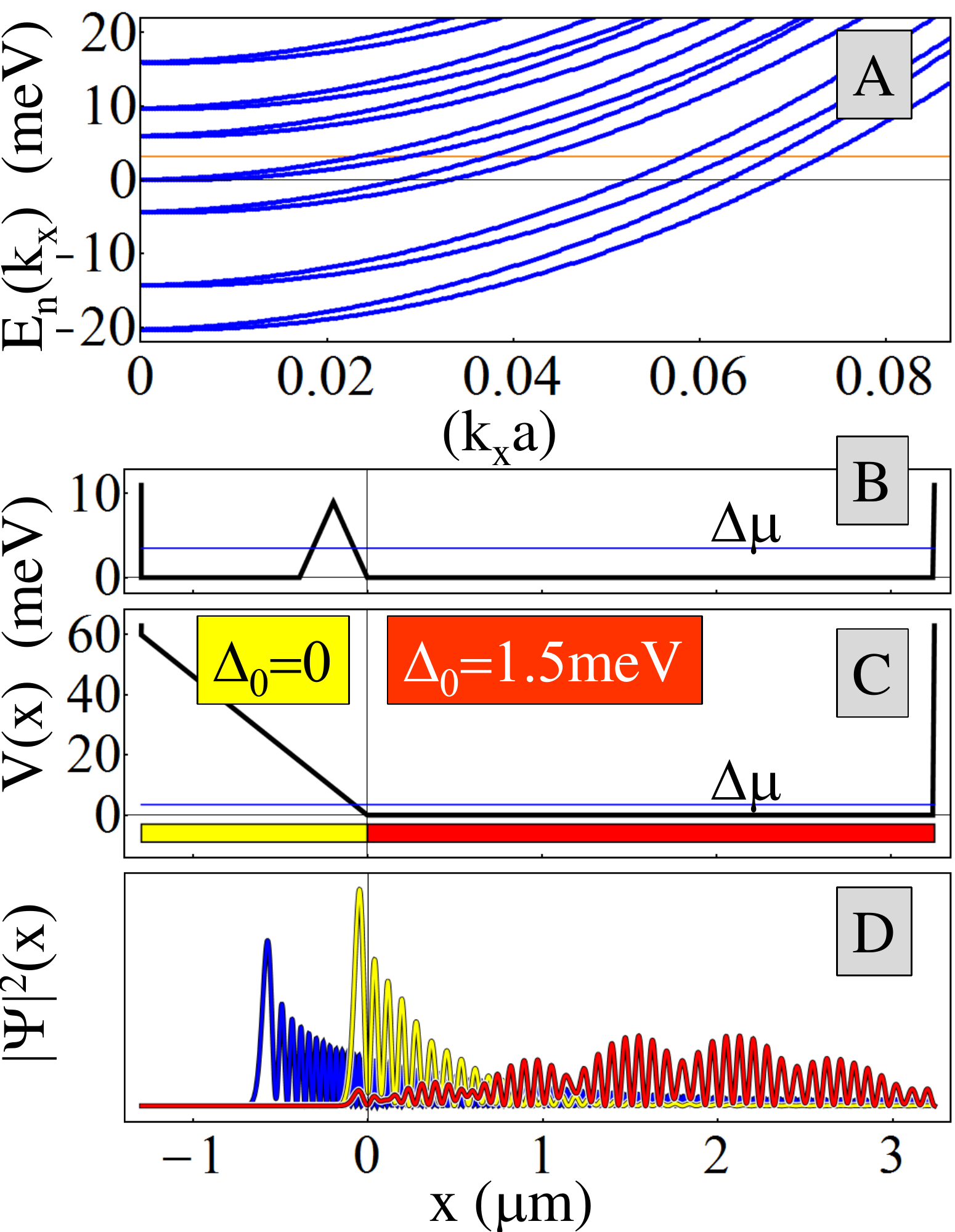}
\vspace{-7mm}
\end{center}
\caption{(Color online) (A) Low-energy spectrum for a non-superconducting nanowire with cross section $L_y\times L_z = 100\times 84$nm. The thin black line corresponds to $\Delta\mu=0$, i.e., a chemical potential at the bottom of the forth band, $(n_y, n_z)=(2,2)$, while the orange (light gray) line is for $\Delta\mu=3.5$meV. (B) and (C) Position dependence of the confining potential $V(x)$. $V(x)$ is assumed to be constant along  the segment of the nanowire covered by the superconductor (red/dark gray).  (D) Typical low-energy states for a system with $\Delta\mu=3.5$meV, zero magnetic field, and $V(x)$ as shown in panel (C). The localized states coming from low-energy bands (blue/dark gray) penetrate deep into the confining barrier. The top occupied band has both localized (yellow/light gray) and delocalized (red/gray) low-energy contributions.  This picture does not change qualitatively at finite magnetic fields.}
\vspace{-6mm}
\label{Fig1}
\end{figure}

We consider a 1D nanowire with four occupied bands and two different values of the chemical potential, one close to the minimum of the top band and another one  that cuts both sub-bands of the top-most band  (see Fig. \ref{Fig1}A). A segment of the nanowire is coupled to a superconductor and separated by a potential barrier from the rest of the wire.
In Fig.~1(B) and 1(C) we show two models of a wire with soft confinement that generate qualitatively similar results. 
The finite barrier shown in Fig. \ref{Fig1}(B) allows the wave functions associated with certain low-energy states to penetrate into the normal segment of the nanowire. In the presence of metallic contacts, these states are strongly hybridized with the states in the lead, which results in a large broadening of the corresponding energy levels. 
Typical low-energy Bogoliubov-de Gennes (BdG) wave-functions for a system in the topologically trivial phase with $\Delta\mu \gg \Delta$ are shown in Fig. \ref{Fig1}(D). The states from low-energy bands penetrate deeper inside the potential barrier, while states from the top band can be either localized near the end or extended throughout the wire. These features remain qualitatively the same in the relevant range of applied Zeeman fields. 

\begin{figure}[tbp]
\begin{center}
\includegraphics[width=0.48\textwidth]{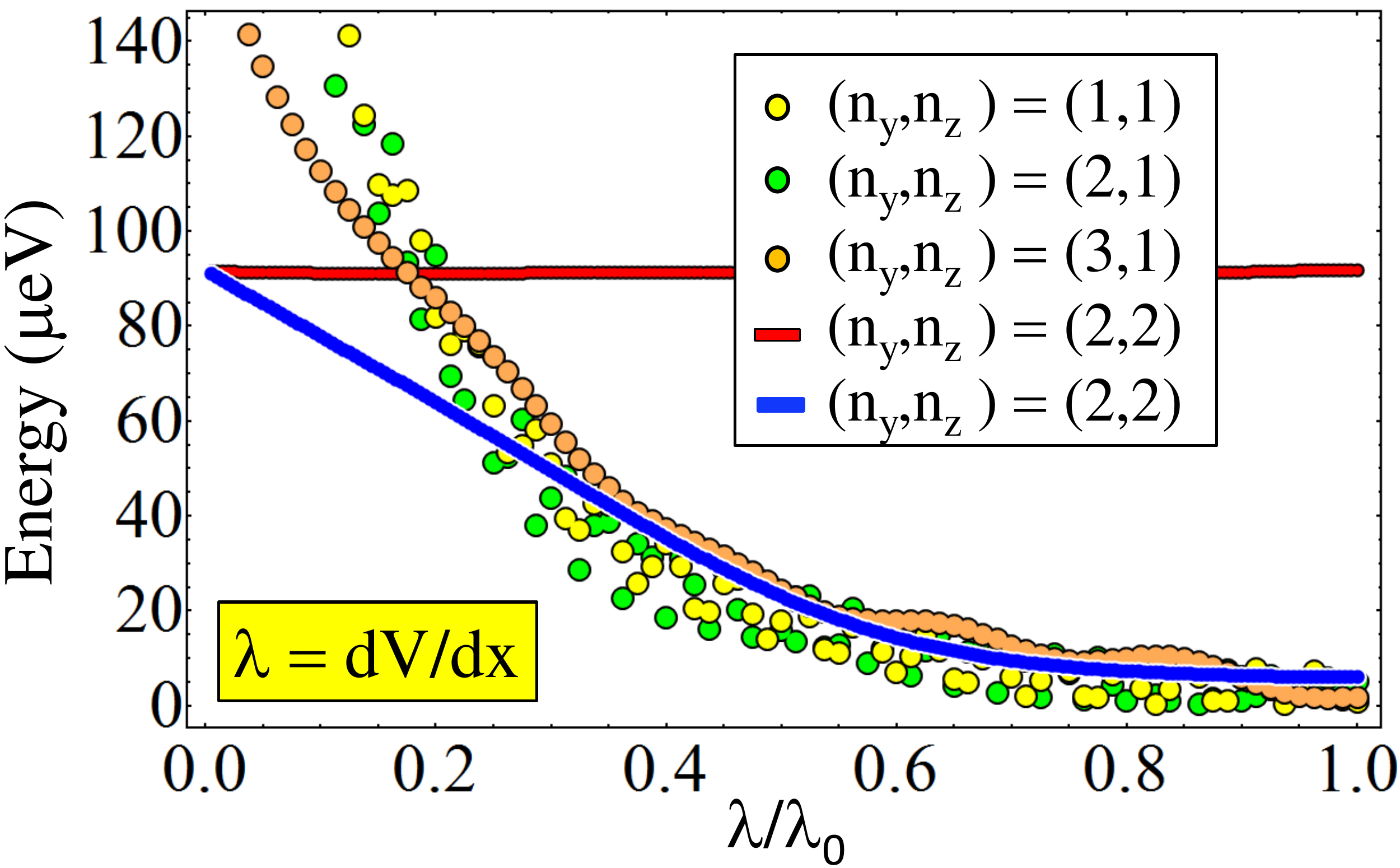}
\vspace{-7mm}
\end{center}
\caption{(Color online) Dependence of the low-energy BdG eigenvalues on the slope of the the smooth confining potential for a system with  $L_x\approx 3\mu$m, $\Delta\mu=3.5$meV, $\Gamma=0.65$meV, and  $V(x)$ as shown in Fig. \ref{Fig1}C. The slope of the confining potential is $\lambda=dV/dx$ and $\lambda_0\approx 23$meV/$\mu$m. The approximately constant contribution from the top band (horizontal red line) corresponds to a state localized near the right end of the wire, which has a hard-wall confinement.}
\vspace{-6mm}
\label{Fig2}
\end{figure}

In Fig.~ \ref{Fig2}, we show the dependence of the low--energy BdG eigenvalues on the slope of the confining potential for a system with $\mu \gg \Gamma \gg \Delta$, which is the condition for the occurrence of non--Majorana near--zero--energy end states in a SM wire with smooth confinement\cite{Kells} . Note that, in this regime, all occupied bands contribute with low-energy states that accumulate near zero-energy as the slope of the potential barrier decreases. 
Recently, we have shown \cite{Stanescu-Tewari} that  for $\Gamma < \Gamma_c$ and $\mu > \mu_c \sim \Delta$  the lowest energy BdG states associated with the top-most band are localized near the wire ends and contribute significantly to the end-of-wire LDOS. For a soft confinement potential, the energies of these localized states gradually decrease with increasing $\Gamma$ and, for $\Gamma \gg \Delta$ (still in the topologically trivial phase, $\Gamma < \Gamma_c$), they become nearly zero energy states. Additional states from the low-energy bands exhibit a similar behavior. Since the states remain localized at the wire ends and contribute significantly to the end-of-wire LDOS as the Zeeman field is varied, the ZBCP arising from the near-zero-energy end states for $\Gamma \gg \Delta$ must necessarily be preceded by a strong dispersion of the LDOS with $\Gamma$, akin to a conventional gap-closing signature. In contrast, for $\mu < \mu_c \sim \Delta$, i.e., when the chemical potential is near the bottom of a certain band, the lowest energy BdG states from the top band  have vanishing amplitude  near the wire ends. Consequently their contribution to the end-of-wire LDOS is negligible and the bulk gap closure at the Majorana TQPT ($\Gamma = \Gamma_c$) is not revealed. Note that, in this regime,  the states associated with low-energy bands have finite energies.  

\begin{figure}[tbp]
\begin{center}
\includegraphics[width=0.5\textwidth]{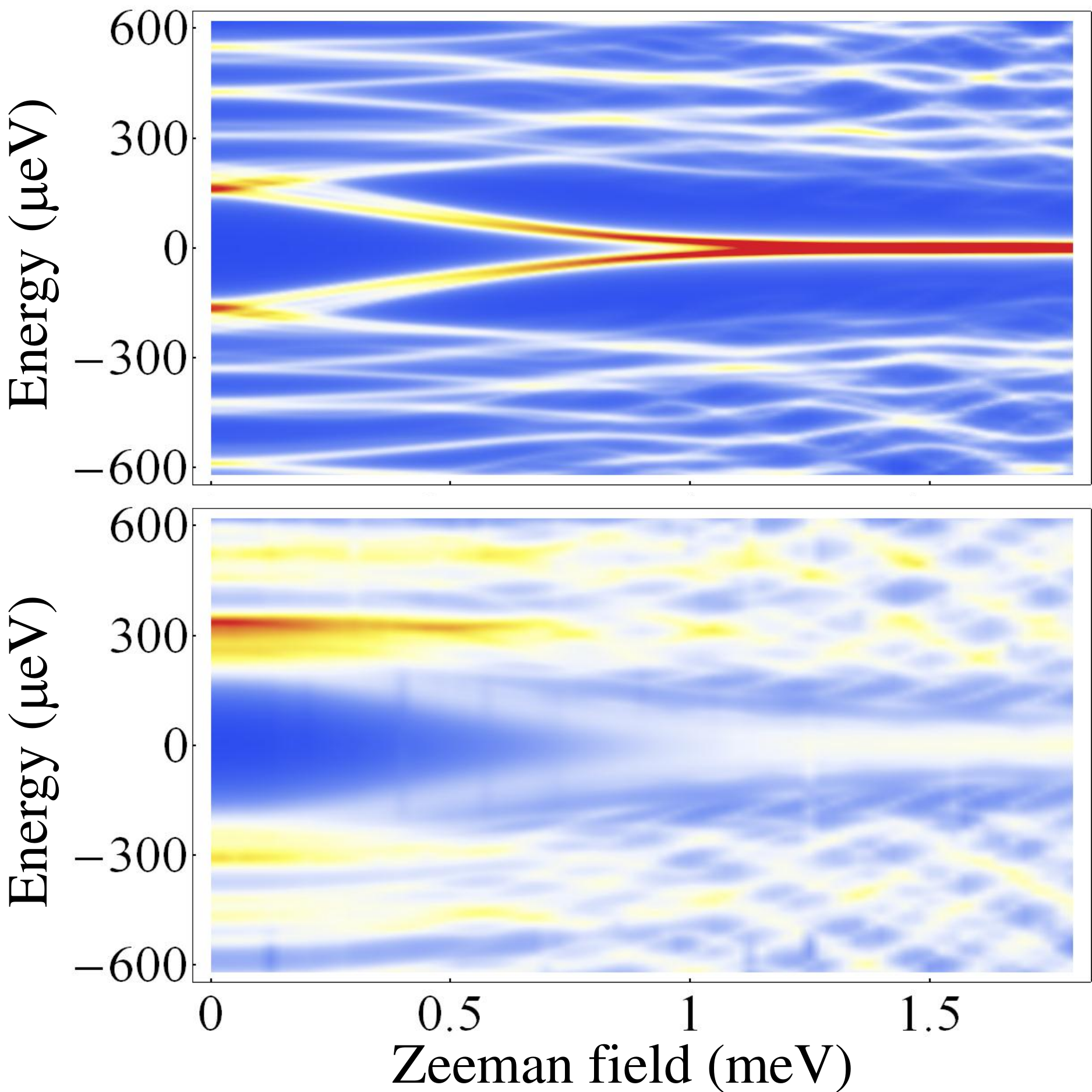}
\vspace{-7mm}
\end{center}
\caption{(Color online) Dependence of the local density of states (LDOS) integrated over the barrier region on the applied magnetic field for a system with $\Delta\mu=3.5$meV (i.e., eight occupied sub-bands) and  $V(x)$ as shown in Fig. \ref{Fig1}C (top panel) and Fig. \ref{Fig1}B (bottom). The energy of the states localized at the soft boundary decreases with $\Gamma$ and a zero-bias peak (ZBP) develops. Note that the closing of the gap is clearly visible, as the spatial dependence of the localized states does not change qualitatively with the magnetic field. For a finite confining barrier (see Fig. \ref{Fig1}B), it is possible that the states corresponding to the low-energy bands penetrate though the barrier and hybridize with states from the leads, which results in a large broadening (see bottom panel).     }
\vspace{-6mm}
\label{Fig3}
\end{figure}

To illustrate the qualitative distinction between the cases $\mu \gg \Gamma \gg \Delta$ (suitable for producing a non-Majorana ZBCP with a soft confinement potential) and $\mu < \Delta$ (suitable for producing MFs when $\Gamma > \sqrt{\Delta^2 + \mu^2}$ ), we first show in Fig.~\ref{Fig3} the dependence of the local density of states integrated over the barrier region on the applied magnetic field for a system with $\mu=3.5$meV (i.e., eight occupied sub-bands). Note that a clear gap closing signature is visible before the emergence of the near ZBCP with increasing $\Gamma$. For a finite barrier corresponding to the confining potential shown in Fig.~1(B), the localized end states  can easily
penetrate to the other side of the barrier and hybridize with the metallic lead. This generates strong broadening, as shown in the bottom panel of Fig. \ref{Fig3}, but does does not change the characteristic features of the LDOS.     
\begin{figure}
\begin{center}
\includegraphics[width=0.48\textwidth]{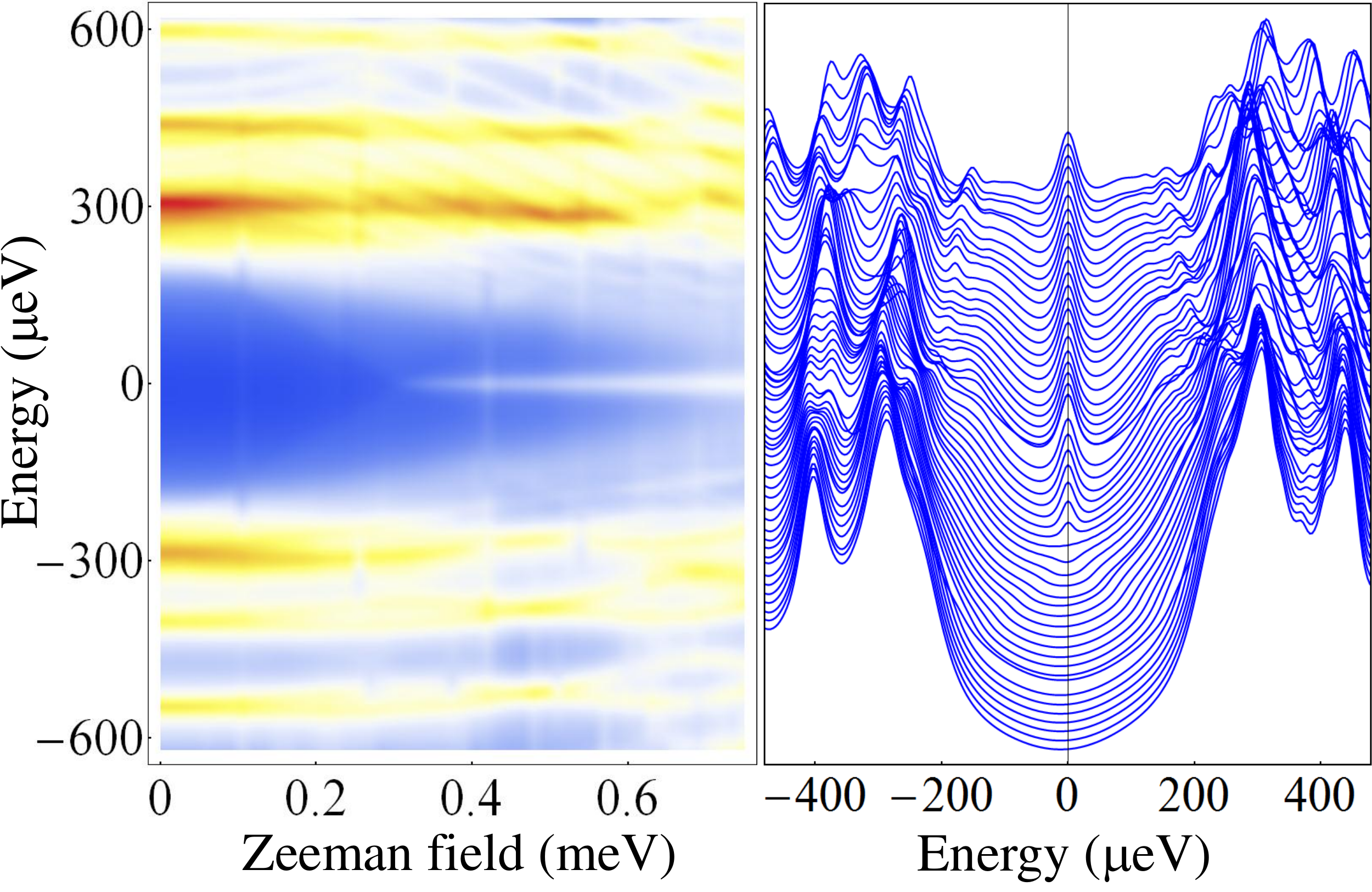}
\vspace{0mm}
\end{center}
\caption{(Color online) Soft superconducting gap as revealed by the LDOS integrated over the barrier region. The chemical potential near the bottom of the forth band ($\Delta\mu=0$ and smooth confinement corresponding to $V(x)$ shown in Fig. \ref{Fig1}B. For $\Gamma>\Gamma_c\approx0.3$meV a ZBP corresponding to the Majorana bound states localized near the finite barrier is clearly visible. Notice the absence of any signature associated with the closing of the quasiparticle gap at the topological quantum phase transition ($\Gamma=\Gamma_c$). The smooth background inside the induced gap $\Delta_{\rm ind}=250\mu$eV is due to contributions from the low-energy states that penetrate though the barrier and hybridize with metallic states from the leads. Constant field cuts (shifted for clarity) are shown in the right panel.}
\vspace{0mm}
\label{Fig4}
\end{figure}
    
Next, we focus on a system with the chemical potential near the bottom of the fourth band, i.e., with $\mu < \mu_c \sim \Delta$. We find that: i) a Majorana ZBCP can be clearly seen for $\Gamma > \Gamma_c$, yet there is no visible gap--closing signature associated with the TQPT, and ii) the SC gap is ``soft'', i.e., the LDOS is characterized by a significant in-gap background at all values of the magnetic field. The results are shown in Fig. \ref{Fig4}. In this case, the 
BdG states from the top band are extended along the nanowire and have a very small amplitude at the ends. Consequently, they do not contribute significantly to the LDOS, which results in the absence of a visible gap-closing signature associated with the TQPT.  For $\Gamma > \Gamma_c$ a ZBCP appears due to a MF state localized at the end of the wire. 
The smooth background inside the induced gap $\Delta_{\rm ind}=250\mu$eV is due to contributions from the low-energy states that penetrate though the barrier and hybridize with metallic states from the leads. In Fig.~\ref{Fig1}(D) we have shown that the BdG states from the lower bands typically have a considerable spectral weight beyond the SC segment of the quantum wire if the potential barrier is finite. These states will hybridize strongly with the metallic states in the lead. We have modeled this effect by introducing a broadening proportional to the spectral weight of the states inside the normal region. 
The resultant LDOS (see Fig. \ref{Fig4}) is characterized by a soft SC gap, a feature present in the experimental data of  Ref.~[\onlinecite{Mourik}]. The key ingredients of the mechanism for the soft gap proposed here are: i) low-energy in--gap states are generated in the presence of a smooth confinement, disorder, etc., ii) a finite potential barrier allows  states with large spectral weight near the end to hybridize with metallic states from the leads, and iii) states from lower--energy bands can penetrate through the barrier and hybridize much stronger that the Majorana mode, which is only weakly broadened and still generates a well defined ZBCP on top of the smooth background. 

\begin{figure}[tbp]
\begin{center}
\includegraphics[width=0.48\textwidth]{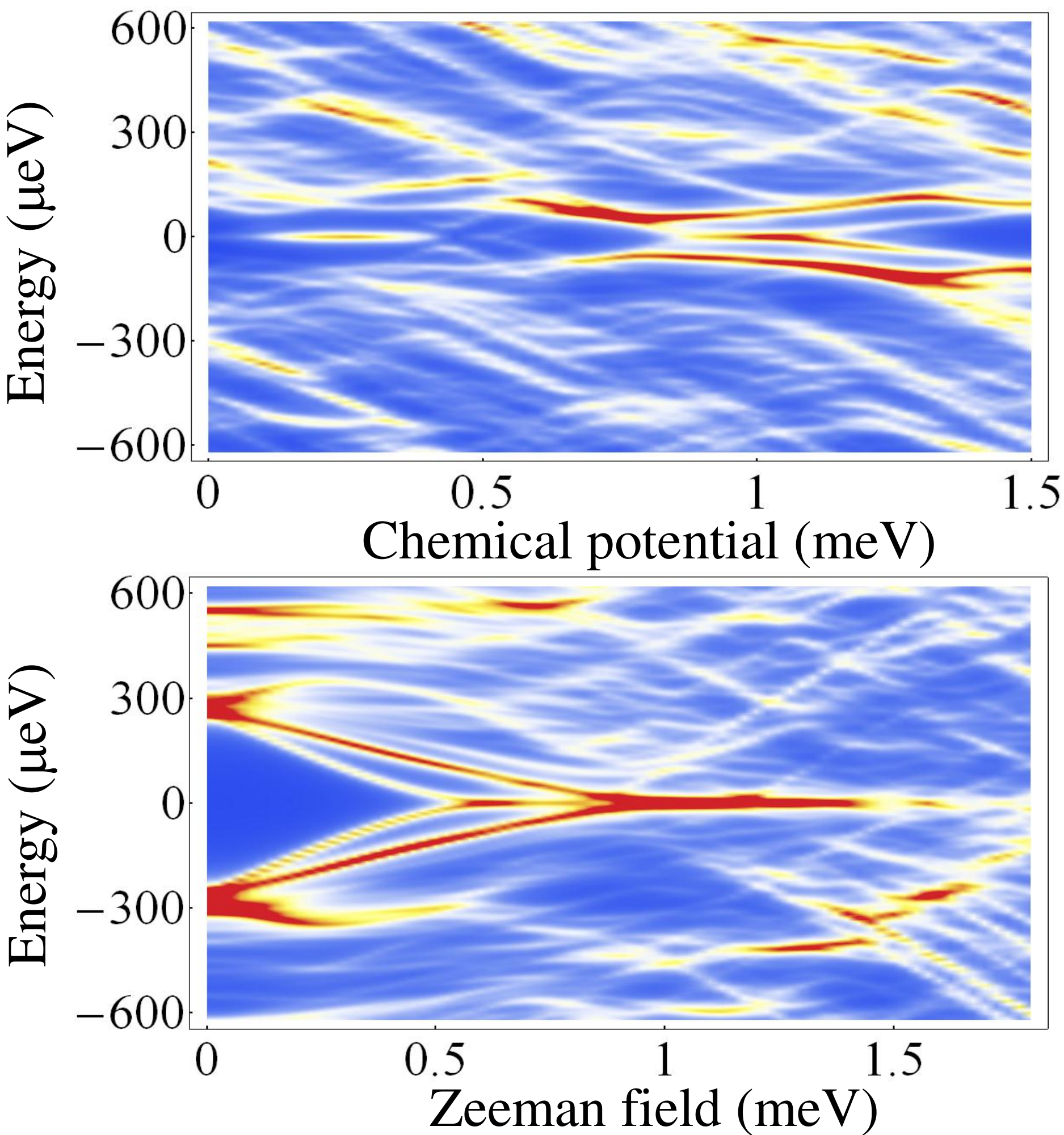}
\vspace{0mm}
\end{center}
\caption{(Color online) Local density of states at the end of the wire in the presence of strong disorder. Top panel: constant magnetic field with  $\Gamma=0.65$meV. Bottom: constant chemical potential $\Delta\mu=1$meV. The closing of the gap associated  with states localized near the wire end is clearly visible in the magnetic field dependence of the LDOS. The disorder potential has constant values inside patches with a characteristic length scale $l_{\rm d}\approx 17$nm. These values vary randomly in the range  $|V_d(x, y)|\leq 3.5$meV. We assume hard-wall confinement.}
\vspace{0mm}
\label{Fig5}
\end{figure}

For zero bias peaks arising in the topologically trivial phase from strong disorder effects we find similar results. In Fig.~5 (top panel) we show
the end-of-wire LDOS as function of the chemical potential for a disordered wire at constant Zeeman field. Near zero bias peaks appear for large values of $\mu$ corresponding to the topologically trivial phase ($\mu$ above $\sim 0.65$ meV). In the bottom panel of Fig.~5 we show the dispersion of the ZBCP with
the Zeeman field for a constant value of $\mu$ in the topologically trivial phase, $\mu=1$meV . Note the appearance of a clear gap closing signature with the Zeeman field before the zero energy peak appears above a threshold magnetic field. The reason for the presence of this signature  is that the states contributing significantly to the end-of-wire LDOS  (and producing the near-zero energy peak above a certain value of $\Gamma$) are localized near the  end of the wire and remain localized near the end even for small values of $\Gamma$ (e.g., for $\Gamma = 0$). With increasing Zeeman field their energies go down but their spectral weight contribution to the end-of-wire LDOS remains nearly the same. Without the clear gap closing signature as a function of $\Gamma$ that precedes the zero energy peak, a ZBCP is unlikely to be due to strong disorder effects.

In conclusion, we have shown that the emergence of a nearly--ZBCP at Zeeman fields corresponding to the topologically trivial phase is necessarily accompanied by a gap closing signature in the end--of--wire LDOS. The absence of such a signature, as observed in the recent experiments \cite{Mourik}, is inconsistent with a nearly--ZBCP due to conventional states \cite{Kells,Liu,Bagrets}  and is only possible before the emergence of a ZBCP generated by localized Majorana fermions \cite{Stanescu-Tewari}. Finally, we have shown that, in a multiband wire with finite confining potential, states from low--energy bands can hybridize strongly with metallic states from the leads, thus generating a soft SC gap.

S. T. thanks DARPAMTO, Grant No. FA9550-10-1-0497 and NSF, Grant No. PHY-1104527 for support.


\end{document}